\documentclass [12pt] {article}
\usepackage{subeqn}
\begin{document}

\setlength\arraycolsep{2pt}

\title{Dyonic  \mbox{Kerr-Newman} black holes, complex scalar field and Cosmic Censorship}
\author{{\.{I}brahim Semiz \thanks{mail: ibrahim.semiz@boun.edu.tr}}\\    
   Department of Physics  \\
   Bo\u{g}azi\c{c}i University\\
Bebek, \.{I}stanbul, TURKEY\\}

\date{ }

\maketitle 

\begin{abstract}
We construct a gedanken experiment, in which a weak wave packet of the complex
massive scalar field interacts with a \mbox{four-parameter} \mbox{(mass,} angular momentum, electric and magnetic \mbox{charges)}  \mbox{Kerr-Newman} black hole. We show that this interaction cannot convert an extreme the black hole into a naked sigularity for any black hole parameters and any generic wave packet configuration. The analysis therefore provides support for the weak cosmic censorship conjecture.
\end{abstract}

\section{Introduction}

A black hole is an object surrounded by an event horizon, i.e. a surface with the property that observers in the exterior cannot receive signals from events in the interior. This is the essential feature of black holes, i.e. what makes them ``black". 

Black holes also harbor spacetime singularities; in some sense, this is what makes them "holes". The  singularity of any black hole is inside ---more precisely, to the future of--- its event horizon.

If a spacetime features a singularity that is not hidden behind an event horizon, i.e. far-away observers can receive signals from it, the singularity is said to be ``naked''. Since initial conditions cannot be specified at a singularity \mbox{---put} another way, anything can come out of the  \mbox{singularity---} a naked singularity would prevent predictability in a spacetime. Therefore, it is conjectured that all singularities are hidden behind event horizons; more precisely, ``naked singularities'' cannot be produced from regular initial conditions and reasonable matter properties \mbox{--- the } ``cosmic censorship'' conjecture (CCC) of Penrose~\cite{penrose.orig.ccc}. This form was later renamed as the  {\em weak}  cosmic censorship conjecture (WCCC)\footnote{Penrose later introduced the {\em strong} CCC, that all singularities should be spacelike or null \cite{penrose.strong.ccc}. The terminology is somewhat misleading, however, since neither statement implies the other. In this work, we do not deal with the strong CCC.}. 

In the absence of a general proof, \mbox{gedanken-} and numerical experiments have been devised  to check the validity of the cosmic censorship conjecture (CCC) under
different limited circumstances, by studying the evolution of various initially regular systems to see if a naked singularity develops \cite{ccc.rev.1,ccc.rev.2,ccc.rev.3}. Black holes are obvious subjects for such investigations.

According to the \mbox{well-known} \mbox{``no-hair''} theorem \cite{mazur,bunting},  of classical General Relativity, stationary black holes in asymptotically flat space can be described uniquely by three variables (\mbox{Mass $M$,} \mbox{charge $Q$} and angular momentum per unit mass, $a$), hence by the  \mbox{Kerr-Newman} metric \cite{nohair}. This spacetime has a horizon  \mbox{---therefore} describes a black  \mbox{hole---} if and only if 
\begin{equation}
M^{2} \geq Q^{2}+a^{2}. \label {criterion}
\end{equation}
If (\ref{criterion}) is not satisfied, the metric describes a naked singularity. Therefore one check for the CCC would be to ask if a  \mbox{Kerr-Newman} spacetime can somehow evolve from a form satisfying (\ref{criterion}) to one that does not.
The first such thought experiment was constructed by Wald, who showed~\cite{wald74} that one cannot overcharge and/or overspin an extreme black hole [i.e. one saturating the inequality (\ref{criterion})] by throwing in test point particles with  electric charge and/or spin. 

However, black holes could also have a magnetic charge, if such existed. A particle with electric charge $Q_{e}$ and magnetic charge $Q_{m}$ is called a dyon, and the interest in the possibility of dyonic black holes has grown since magnetic monopoles have been predicted in various extensions of the standard model of particle physics.  The metric of such a black hole is identical to the  \mbox{Kerr-Newman} metric with the replacement $Q^{2}  \rightarrow Q^{2}_{e} + Q^{2}_{m} $, as can be seen by noting that the
 \mbox{energy-momentum} tensor is invariant under the electromagnetic duality transformation, which also gives the electromagnetic \mbox{field/potential} of the dyonic black hole \cite{semiz.1}.  Wald's argument was generalized (for spinless test particles) to the case of the dyonic  black hole by Hiscock \cite{hiscock} and independently, by Semiz  \cite{semiz.1}\footnote{However, \cite{hiscock} claims that the test particle approximation is not valid everywhere in the parameter space, and that inequality (\ref{criterion}), therefore the CCC can be violated, e.g. for the interaction of an electric test particle and an extreme magnetic black hole.}.

The problem of trying to overcharge a maximally charged black hole with another kind of charge was revisited by Bekenstein \& Rosenzweig  \cite{bekenst-rosenzwg}, where nonzero size of the particle  \mbox{---classical} and  \mbox{quantum---} is pointed out and argued to save CCC. Hod \cite{hod} considered lowering the  \mbox{differently-charged} particle slowly towards the horizon (to minimize its energy), and argued that the finite size of the particle helps preserving the horizon in this case too, aided by the polarization of the black hole.

Thought experiments that do not use the electric black  \mbox{hole-magnetic} test particle  combination and  \mbox{duality-rotated} cases include that of Ford \& Roman \cite{ford}, who considered sending a flux of negative energy down an extreme  \mbox{Reissner-Nordstr\"{o}m} (R-N)
 black hole by using a quantized massless scalar field. These negative energy fluxes are allowed in Quantum Field Theory, but the authors find that the exposure of the naked singularity can be only temporary (``Cosmic Flashing") and that the effect probably gets swamped by quantum fluctuations. Jensen  \cite{jensen}, analyzed the effect of the collapse of a spherical  \mbox{domain-wall} with negative energy onto an extreme  \mbox{Reissner-Nordstr\"{o}m} (R-N)
black hole and decided that the WCCC remains valid in this case.  Hubeny \cite{hubeny} argued that one can overcharge a  R-N black hole by starting not from extreme, but slightly  \mbox{sub-extreme} condition, i.e one may ``jump over" the extreme condition. Hod \cite{hod.3} made the same claim for general black holes while pointing out that curved-space modifications to particle self-energy may be relevant. The concept of jumping over the extreme condition was also treated by Jacobson \& Sotiriou \cite{Jacobson-Sot}.
Hod \& Piran \cite{hod-piran} suggested that the WCCC can be violated classically,  but may be saved by Quantum Gravity.  de Felice \& Yu \cite{deFelice-yu} calculated that a naked singularity can be produced by throwing a  \mbox{fast-spinning} flat disk into an extreme  R-N black hole. More recently, Matsas et. al.  \cite{matsas.1,matsas.2} made the same claim for quantum tunneling of low-energy, high-angular-momentum massless scalar particles; and the claim was disputed by Hod \cite{hod.1,hod.2}. Bouhmadi-L\'{o}pez et. al. \cite{hi-dim}  generalized Wald's thought experiment to higher-dimensional black holes and black rings.

It has also been argued \cite{wang-su-et-al} that the extreme black hole, which was the starting point of some of the  \mbox{above-mentioned} works, can not be reached starting from a  \mbox{non-extreme} black hole, which seems to be in accord with the claims \cite{hawking-et-al-95, t'hooft} that the extreme black holes are not the limit of normal black holes; in particular, that the entropy of the extreme black hole is zero, contrary to the  \mbox{Bekenstein-Hawking} formula.

The status of the CCC is currently unresolved. As mentioned above, no general proof exists. Some numerical experiments, e.g.  \cite{numerical}, and some of the  \mbox{above-mentioned} references suggest its violation, but these violations are thought to be \mbox{``non-generic"} or possibly``of measure zero" \cite{ccc.rev.1}, especially since what constitutes ``reasonable matter properties" is not clear.

In the present work, we consider another thought experiment  to check the validity of WCCC: We try to see if a {\em fully general} (dyonic Kerr-Newman) classical extreme black hole can be pushed beyond extremality by absorbing electric charge and/or angular momentum from a  {\em fully general} (complex, massive) classical  scalar {\em test} field.
 
We require the scalar field to have finite energy (e.g. constitute a wave packet). Initially
($t \rightarrow -\infty$), there is just the black hole, and no scalar field.
Then the field comes in from infinity, part of it gets swallowed by the black
hole, part of it gets reflected back to infinity; energy, charge and angular
momentum get transferred. As $t \rightarrow \infty$, the scalar field decays
away, and by the  \mbox{no-hair} theorem, the geometry becomes again a
\mbox{Kerr-Newman} geometry, but with new values of $M$, $Q_{e}$ and $a$.

We calculate the changes of the black hole parameters by first constructing expressions for fluxes of charge, angular momentum and energy carried by the scalar field directly and by its accompanying electromagnetic field, then evaluating them to the lowest nontrivial order in a perturbation expansion around the initial \mbox{Kerr-Newman} configuration. We ask if --in this thought experiment and to this order-- inequality~(\ref{criterion}), therefore the Weak Cosmic Censorship Conjecture can be violated for any
combination of black hole parameters.

\section{Changes in the mass, electric charge and angular momentum of the black hole} \label{sec:ParamChanges}

\indent The  \mbox{Kerr-Newman} metric describing the  \mbox{four-parameter} black hole is given by
\begin{subequations}
\begin{eqnarray}
ds^{2} = g_{\mu\nu}dx^{\mu}dx^{\nu}
       & = & \frac{a^{2}\sin^{2}\theta-\Delta}{\rho^{2}}dt^{2}
+\frac{(r^{2}+a^{2})^{2}-\Delta a^{2}\sin^{2}\theta}{\rho^{2}}\sin^{2}\theta
                                                    d\phi^{2} \nonumber \\
       & + & 2\frac{\Delta-(r^{2}+a^{2})}{\rho^{2}}a\sin^{2}\theta dt d\phi
+\frac{\rho^{2}}{\Delta}dr^{2}+\rho^{2}d\theta^{2}  \label{eq:dknmetric.a}
\end{eqnarray}
where
\begin{eqnarray}
\rho^{2} & = & r^{2}+a^{2}\cos^{2}\theta\\
\Delta & = & r^{2}-2Mr+a^{2}+Q_{e}^{\:\:2}+Q_{m}^{\:\:2} \label{eq:dknmetric.c}
\end{eqnarray}
\end{subequations}
and we are using MTW~\cite{mtw} notation and conventions.

The vector potential describing the electromagnetic field of the black hole is
\begin{subequations}
\begin{eqnarray}
A_{t} & = &  -Q_{e}\frac{r}{\rho^{2}}+Q_{m}\frac{a\cos \theta}{\rho^{2}} \label{eq:pot0} \\
A_{r} & = & A_{\theta}=0 \label{eq:pot1} \\
A_{\phi} & = & Q_{e}\frac{ar\sin^{2}\theta}{\rho^{2}}
+Q_{m}[\pm1-\cos \theta\frac{r^{2}+a^{2}}{\rho^{2}}] \label{eq:pot3}
\end{eqnarray}
\end{subequations}

The two signs in the last term correspond to the two gauges used to avoid the ``Dirac string" in the magnetic part of the vector potential, as described in \cite{semiz.1}. A wavefunction in this formalism is called a section \cite{semiz.2, wu-yang}.

To calculate the changes in the black hole parameters, we start from the associated conservation laws. As is well known, the energy and angular momentum conservation laws are consequences of spacetime symmetries and locally take the form of continuity equations
\begin{equation}
(T^{\mu \nu} X_{\nu})_{;\mu} = 0 \label{eq:cons}
\end{equation}
where $T^{\mu \nu}$ is the  \mbox{energy-momentum}  tensor for test particles or fields
and $X^{\nu}$ is the Killing vector generating the symmetry.

We also have the current continuity equation
\begin{equation}
(\sqrt{-g} \, j^{\nu})_{,\nu} = 0 \label{eq:cons.current}
\end{equation}
where $j^{\nu}$ is the current density four-vector.

The \mbox{Kerr-Newman} metric is independent of $t$ and $\phi$, therefore its
Killing vectors are $\frac{\partial}{\partial x^{0}}$ and $\frac{\partial} {\partial x^{3}}$.
For the first of these, integrating (\ref{eq:cons}) over a spacelike volume we get
\begin{equation}
\frac{d}{dx^{0}} \int_{V} (\sqrt{-g} \, T_{0}^{\;\;0})\; d^{3} x
= - \int_{V} (\sqrt{-g} \, T_{0}^{\;\;i})_{,i} \; d^{3} x
= - \int_{S} \sqrt{-g} \, T_{0}^{\;\;i} dS_{i} \label{eq:cons.exp}
\end{equation}
where $S$ is the surface (boundary) of the volume $V$. 

The mass of a physical system is defined only in the asymptotically flat part of space at infinity, if it 
exists~\cite[\S 19.3, \S 20.2]{mtw}; and the energy density is $-T_{0}^{\;\;0}$~\cite[eq.(5.11)]{mtw}. Therefore we have
\begin{equation}
\left(\frac{dM}{dt}\right)_{\rm b.h} =  \int_{S_{\infty}} \sqrt{-g} \, T_{0}^{\;\;1} d\theta d\phi
                                                   \label{eq:dm/dt}
\end{equation}
where the label b.h. stands for black hole, $S_{\infty}$ is the spherical
surface as $r \rightarrow \infty$ (cf. \cite{mtw}, eq.(20.27)). Similarly, eqs. (\ref{eq:cons}) and (\ref{eq:cons.current}) lead to
\begin{eqnarray}
\left(\frac{dL}{dt}\right)_{\rm b.h}
& = & - \int_{S_{\infty}} \sqrt{-g} \, T_{3}^{\;\;1} d\theta d\phi
                                                   \label{eq:dl/dt}    \\
\left(\frac{dQ_{e}}{dt}\right)_{\rm b.h}
& = & - \int_{S_{\infty}} \sqrt{-g} \, j^{1} d\theta d\phi
                                                   \label{eq:dq/dt}
\end{eqnarray}

To evaluate $T_{\mu \nu}$ for our problem, we start from the Lagrangian of
the combined scalar and electromagnetic fields :
\begin{equation}
{\cal L} = \frac{1}{8 \pi} [ -g^{\alpha \beta}
(\partial_{\alpha} - i e A_{\alpha}) \psi^{*}
(\partial_{\beta} + i e A_{\beta}) \psi - \mu^{2} \psi^{*} \psi]
- \frac{1}{16 \pi} F_{\alpha \beta} F^{\alpha \beta}  \label{eq:lagrangian}
\end{equation}
The \mbox{energy-momentum} tensor is defined by \cite[\S 21.3]{mtw}
\begin{equation}
T_{\mu \nu}
= -2 \frac{\delta {\cal L}}{\delta g^{\mu \nu}} + g_{\mu \nu} {\cal L}
\end{equation}
hence
\begin{equation}
T_{\mu}^{\;\;\nu} = \frac{1}{8 \pi} [(\partial_{\mu} - i e A_{\mu}) \psi^{*}
(\partial^{\nu} + i e A^{\nu}) \psi + (\partial^{\nu} - i e A^{\nu}) \psi^{*}
(\partial_{\mu} + i e A_{\mu}) \psi]  \\
+ \frac{1}{4 \pi} F_{\mu \beta} F^{\nu \beta} + \delta_{\mu}^{\;\; \nu} {\cal L}.
                                 \label{eq:tmunu}
\end{equation}

Also, the current density $j^{1}$ is given by 
\begin{equation}
j^{1} = \frac{ie}{8 \pi} [\psi^{*} (\partial^{1} + i e A^{1}) \psi
           - \psi (\partial^{1} - i e A^{1}) \psi^{*}]. \label{eq:j1}
\end{equation}

We assume deviations from background are small, (`test field') and expand the fields in some
small parameter $\epsilon$. The background is the dyonic
\mbox{Kerr-Newman} spacetime, therefore for the zeroth order fields we have $\psi^{(0)} = 0$,
$F_{\mu\nu}^{(0)} \neq 0 \Longrightarrow A_{\mu}^{(0)} \neq 0$.

\indent From the Lagrangian (\ref{eq:lagrangian}) we derive the equations of motion for
the fields. The equation of motion for $\Psi$ gives  to first order (the zeroth order part has no content, $0=0$)
\begin{equation}
\frac{1}{\sqrt{-g}}(\partial_{\mu} + ie A^{(0)}_{\mu})
                  [\sqrt{-g}(\partial^{\mu} + ie A^{(0) \mu})] \psi^{(1)}
                  = \mu^{2} \psi^{(1)}       \label{eq:eompsi.1}
\end{equation}
Similarly, the equation of motion for $F^{\mu \nu}$ gives to zeroth and first orders
\begin{equation}
F^{(0) \alpha \beta}_{\;\;\;\;\;\;\;\; ;\beta} = 0 ,  \;\;\;\;\;\;\;\;   F^{(1) \alpha \beta}_{\;\;\;\;\;\;\;\; ;\beta} = 0 \label{eq:eom.f.01}
\end{equation}
and to second order
\begin{equation}
F^{(2) \alpha \beta}_{\;\;\;\;\;\;\;\; ;\beta} = 4 \pi \left\{\frac{ie}{8 \pi}
[\psi^{*(1)} (\partial^{\alpha} + ie A^{(0)\alpha}) \psi^{(1)}
 - \psi^{(1)} (\partial^{\alpha} - ie A^{(0)\alpha}) \psi^{*(1)}]\right\}
= 4 \pi j^{(2) \alpha}   \label{eq:eom.f.2}
\end{equation}

We require there to be no photons. Since there is no source for the
electromagnetic field to zeroth or first order, this
requirement, together with boundary conditions at infinity,
makes $F^{(0) \alpha \beta} + F^{(1) \alpha \beta}$ the dyonic
\mbox{Kerr-Newman} solution. Since $F^{(0) \alpha \beta}$ is the background,
and therefore already is \mbox{Kerr-Newman}, we have
$F^{(1) \alpha \beta} = 0 \Longrightarrow A^{(1)}_{\mu} = 0$
and $F^{(2) \alpha \beta}_{\;\;\;\;\;\;\;\; ;\beta} = 4 \pi j^{(2)\alpha}$,
with $j^{(2)\alpha}$ is made up of $\psi^{(1)}$'s, as above.

The  \mbox{lowest-order} contributions to $T_{0}^{\;\;1}$, $T_{3}^{\;\;1}$, and $j^{1}$
are second order \mbox{in $\epsilon$:} Terms with $\psi$, $\psi^{*}$ and their
derivatives are obvious. $FF$ terms also vanish to zeroth order, a result one
can verify by calculation and understand physically: There are no energy, or
angular momentum flows on the exact \mbox{Kerr-Newman} metric.
$FF$ terms also vanish
to first order, because $F^{(1) \alpha \beta} = 0$. Therefore
{\em the lowest order contributions to $dM/dt$, $dL/dt$, and $dQ_{e}/dt$} [given in eqs.(\ref{eq:dm/dt}-\ref{eq:dq/dt})] {\em are second order in $\epsilon$.}

\section{Treating separated equations} \label{treateqn}

\indent As shown in \cite{semiz.2}, equation
(\ref{eq:eompsi.1}) can be separated in  \mbox{Boyer-Lindquist} coordinates by
\begin{equation}
\psi^{(1)} =  R(r) \Theta(\theta) e^{-i\omega t}
                                e^{i(m \mp e Q_{m}) \phi}. \label{eq:psisep}
\end{equation}
The angular equation
\begin{equation}
\frac{1}{\sin \theta}\frac{d}{d \theta}
            \left(\sin \theta \frac{d \Theta}{d \theta}  \right)
+ \left[ a^{2} (\omega^{2}-\mu^{2}) \cos^{2}\theta
        - \frac{(m-eQ_{m}\cos\theta)^{2}}{\sin^{2}\theta}
        - 2 a \omega e Q_{m} \cos\theta + \lambda \right] \Theta = 0 \nonumber
\end{equation}
constitutes a  \mbox{Sturm-Liouville} problem, therefore the solutions form a complete set and can be orthonormalized. Labeling the eigenvalue $\lambda$ by $l$, we choose
\begin{equation}
\int Y_{ql'm'}^{*}(a,\mu,\omega,\theta,\phi)
          Y_{qlm}(a,\mu,\omega,\theta,\phi) d\Omega
= \delta_{ll'} \delta_{mm'}    \label{eq:y.norm}
\end{equation}
where
\begin{equation}
Y_{qlm}(a,\mu,\omega,\theta,\phi) = \Theta_{qlm}(a,\mu,\omega,\theta)
                                           e^{i(m \mp e Q_{m})\phi}
\end{equation}
are functions we had decided to call {\em monopole
spheroidal harmonics} \cite{semiz.2}.

The radial equation is
\begin{equation}
\frac{d}{d r} \left( \Delta \frac{d R}{d r}  \right)
+ \left\{ \frac{1}{\Delta} [(r^{2}+a^{2})\omega + eQ_{e}r -am ]^{2}
          - \mu^{2}r^{2} + 2 am\omega - a^{2}\omega^{2} - \lambda
                                  \right\} R = 0    \label{eq:radial}
\end{equation}
Switching the independent variable to the `tortoise' coordinate $r^{*}$
defined by
\begin{equation}
\frac{dr^{*}}{dr} = \frac{r^{2}+a^{2}}{\Delta} \Rightarrow
\Delta \frac{d}{dr} = (r^{2}+a^{2})\frac{d}{dr^{*}}
\end{equation}
and substituting
\begin{equation}
R(r) = \frac{U(r)}{\sqrt{r^{2}+a^{2}}}
\end{equation}
for the dependent variable,
we get after extensive manipulation
\begin{eqnarray}
\lefteqn{\frac{d^{2}}{dr^{* 2}} U+ \left\{
\left[\frac{(r^{2}+a^{2})\omega + eQ_{e}r - am}{r^{2}+a^{2}} \right]^{2}\right.
                          }  \nonumber  \\
& & \left.
- \frac{\Delta (\mu^{2}r^{2} - 2 am\omega + a^{2}\omega^{2} + \lambda)}
      {(r^{2}+a^{2})^{2}} - \frac{\Delta^{2} (a^{2}-2r^{2})}{(r^{2}+a^{2})^{4}}
- \frac{2 r \Delta (r-M)}{(r^{2}+a^{2})^{3}}  \right\} U = 0 \nonumber  \\
\label{eq:radial.u}
\end{eqnarray}
which near the horizon ($r \rightarrow r_{+}$, $r^{*} \rightarrow -\infty$)
reduces to (easily seen, since $r_{+}$ is the larger root of $\Delta$)
\begin{equation}
\frac{d^{2} U}{dr^{* 2}}
+ \left[\frac{(r_{+}^{2}+a^{2})\omega + eQ_{e}r_{+} - am}
             {r_{+}^{2}+a^{2}} \right]^{2} U = 0
= \frac{d^{2} U}{dr^{* 2}} + \bar{\omega}^{2} U = 0  \label{eq:rad.u.h}
\end{equation}
(where $\bar{\omega}$ has been defined in the equation) and near infinity (as $r^{*} \rightarrow r \rightarrow \infty$)
becomes
\begin{equation}
\frac{d^{2} U}{dr^{* 2}} + (\omega^{2}-\mu^{2}) U = 0  \label{eq:rad.u.inf}
\end{equation}
That is, in terms of $U$ and $r^{*}$, the radial problem at a given frequency looks like a \mbox{one-dimensional} Schr\"{o}dinger scattering problem with two (different)
constant potentials at two ends and a complicated potential well in the middle.

The solutions for $U$ near the horizon are, from eq.(\ref{eq:rad.u.h}),
\begin{equation}
U = B e ^{\pm i \bar{\omega} r^{*}}
\end{equation}
and near infinity, from eq.(\ref{eq:rad.u.inf}),
\begin{equation}
U = e ^{\pm i k r^{*}} \;\; {\rm for} \;\; \omega^{2} > \mu^{2} \; , \;
                  {\rm where} \;\;  k^2 = \omega^{2} - \mu^{2}
\end{equation}
and
\begin{equation}
U = e^{\pm \kappa r^{*}} \;\; {\rm for} \;\; \omega^{2} < \mu^{2}  \; , \;
      {\rm where} \;\; \kappa^2 = \mu^{2} - \omega^{2} \;\;
\end{equation}

For $\omega^{2} < \mu^{2}$, the exponentially growing solution
$e^{\kappa r^{*}}$ is unphysical, and the
solution $e^{-\kappa r^{*}}$ corresponds to a bound particle and vanishes
exponentially at infinity, and therefore will not contribute to our integrals
for $dM/dt$, $dQ/dt$, $dL/dt$.

For $\omega^{2} > \mu^{2}$, the outgoing solution is unphysical at the future
event horizon: It varies infinitely fast as seen by an ingoing,
\mbox{future-directed} observer. We are left with the ingoing solution:
\begin{equation}
\lim_{r \rightarrow r_{+}} U_{\omega l m}(r^{*})
                              = B_{\omega l m} e^{-i \bar{\omega}r^{*}}
\end{equation}
and for the same mode,
\begin{equation}
\lim_{r \rightarrow \infty} U_{\omega l m}(r^{*})
                          = e^{-i k r^{*}} + A_{\omega l m} e^{i k r^{*}}
\end{equation}
where the solution has been normalized such that the coefficient of the
first term on the \mbox{right-hand} side of the last equation is unity.
In other words, this corresponds to a wave of unit amplitude coming in from
infinity, being transmitted into the black hole with amplitude $B$ and
reflected back to infinity with amplitude $A$.

Since eq.(\ref{eq:radial.u}) is real, the complex conjugate of any solution
is also a solution, and since (\ref{eq:radial.u})
has no first derivative, the Wronskian of the two solutions $U$ and $U^{*}$
\begin{equation}
W = U \frac{dU^{*}}{dr^{*}} - U^{*} \frac{dU}{dr^{*}}
\end{equation}
is constant; therefore its values at the two limits can be put equal to each
other, giving after a few lines
\begin{equation}
\lim_{r \rightarrow r_{+}} W = \bar{\omega} B B^{*} =
\lim_{r \rightarrow \infty} W = k (1-A A^{*})  \label{eq:wronski}
\end{equation}
where the labels $\omega l m$ are implied for the relevant variables
in the last two equations.

\section{Testing the Cosmic Censorship Conjecture} \label{sec:KG.criterion}

\indent To test the inequality (\ref{criterion}),
let us define a cosmic censorship indicator,
\begin{equation}
CCC =  M^{2} - (Q^{2} + a^{2})
\end{equation}

Then, using $dQ_{m}/dt=0$ and $a=L/M$, and implicitly defining $dC/dt$,
\begin{equation}
\delta (CCC)  =   \int \frac{dC}{dt}  dt  =  \int \frac{2}{M} \left\{ (M^{2}+a^{2}) \frac{dM}{dt} - M Q_{e} \frac{dQ_{e}}{dt} - a \frac{dL}{dt} \right\} dt.
\end{equation}

Recalling the results of section \ref{sec:ParamChanges}, we have to second order in $\epsilon$
\begin{equation}
 \frac{dC}{dt} = \int_{S_{\infty}} \sqrt{-g} [ (M^{2}+a^{2}) T_{0}^{\;\;1} + MQ_{e} j^{1}
              + a T_{3}^{\;\;1} ] d \theta \; d \phi \label{eq:dC/dt1}
\end{equation}

In the pure electromagnetic contribution to $dC/dt$ (the part coming from the $F_{\mu \beta} F^{\nu \beta}$ term in $T_{\mu}^{\;\; \nu}$, see eq.(\ref{eq:tmunu})) we write $F_{\mu \beta}$ in terms of $A_{\mu}$:
\begin{equation}
\left( \frac{dC}{dt}\right)_{\rm em} = \frac{1}{2 \pi M} \int_{S_{\infty}} \sqrt{-g} \left\{  \right.
 (M^{2}+a^{2})  (A_{\alpha,0} - A_{0,\alpha}) F^{1 \alpha}  \nonumber \\
 + a  \left. [A_{\alpha , 3} - (A_{3} \mp Q_{m} )_{,\alpha}]  \; F^{1 \alpha}
\right\} d \theta \; d \phi 
\end{equation}
where the $ \mp Q_{m}$ was added to $A_{3}$ for later convenience. Rewriting the  \mbox{$\alpha$-derivative} terms as integrals by part,we get
\begin{eqnarray}
\left( \frac{dC}{dt}\right)_{\rm em} & = & \frac{1}{2 \pi M}  \int_{S_{\infty}} \left\{ 
 (M^{2}+a^{2}) [\sqrt{-g} A_{\alpha,0} F^{1 \alpha} - (\sqrt{-g} A_{0} F^{1 \alpha})_{,\alpha}
 +  A_{0} (\sqrt{-g}  F^{1 \alpha})_{,\alpha} ] \right. \nonumber \\
& + &  \left. a \left[ \sqrt{-g}  A_{\alpha , 3}  F^{1 \alpha}  - [(A_{3} \mp Q_{m}) \sqrt{-g} \; F^{1 \alpha}]_{,\alpha}
 +  (A_{3} \mp Q_{m}) (\sqrt{-g} \; F^{1 \alpha})_{,\alpha}\right] \right\} 
 d \theta \; d \phi \hspace{0.8 cm}
\end{eqnarray}
In the second and fifth terms, the $\alpha=2$ parts vanish when integrated over $\theta$,
because of $\sin\theta$ in 
\begin{equation}
\lim_{r \rightarrow \infty} \sqrt{-g} = r^{2} \sin \theta,    \label{rhosqlimit}
\end{equation}
 the $\alpha=3$ parts vanish when integrated over $\phi$, because $\phi$ is cyclic. In the third and sixth terms, $(\sqrt{-g} \; F^{1 \alpha})_{,\alpha}=4 \pi \sqrt{-g} \; j^{1}$. Also, the lowest nonvanishing contributions to $A_{\alpha,0}$, $A_{\alpha,3}$ and $j^{1}$ are of second order, so we get for the pure electromagnetic contribution to $\delta (CCC)$
\begin{eqnarray}
\delta (CCC)_{\rm em} = \frac{1}{2 \pi M}  \int^{\infty}_{-\infty} \int_{S_{\infty}} \left\{ 
 (M^{2}+a^{2}) [\sqrt{-g} F^{(0)1\alpha} A^{(2)}_{\alpha,0} - (\sqrt{-g} A_{0} F^{10})_{,0}
 + 4\pi A^{(0)}_{0} \sqrt{-g} j^{(2)1} ] \right.  \nonumber \\ 
 + a \left[ \sqrt{-g}  F^{(0)1 \alpha} A^{(2)}_{\alpha , 3}   - [(A_{3} \mp Q_{m}) \sqrt{-g} \; F^{10}]_{,0}
\left. +  4 \pi \sqrt{-g} (A^{(0)}_{3} \mp Q_{m}) j^{(2)1} \right] \right\} 
 d \theta \; d \phi \; dt   \hspace{0.5 cm}
 \label{eq:dCCC1}
\end{eqnarray}
$F^{(0)1 \alpha}$ is a part of the background field, therefore independent of time, so the time integral can be done for the first term. The second and fifth terms are now total time derivatives, so those time integrals can be done also. On $S_{\infty}$, $A^{(0)}_{0}$ goes to $-Q_{e}/r$, therefore becomes  \mbox{angle-independent,} so the angular integral can be done in the third term, giving $dQ_{e}/dt$ via 
eq.(\ref{eq:dq/dt}). $F^{(0)1 \alpha}$ is also  \mbox{$\phi$-independent,} so the  \mbox{$\phi$-integral} in the fourth term applies to $A^{(2)}_{\alpha , 3}$ and vanishes because $\phi$ is cyclic. So $\delta (CCC)_{\rm em}$ becomes
\begin{eqnarray}
\frac{1}{2 \pi M} \left\{  \right. 
& \int_{S_{\infty}}&   \sqrt{-g} (M^{2}+a^{2}) F^{(0)1 \alpha}
A^{(2)}_{\alpha} |^{t = \infty}_{t = -\infty} d \theta \; d \phi 
-  \int_{S_{\infty}} 
(\sqrt{-g}  (M^{2}+a^{2}) A_{0} F^{10}) |^{t = \infty}_{t = -\infty} d \theta \; d \phi \nonumber \\
&-&  4 \pi  (M^{2}+a^{2}) \frac{Q_{e}}{r} \int_{-\infty}^{+\infty} \frac{dQ_{e}}{dt} dt  
-  \int_{S_{\infty}}  \left. \left. a
[\sqrt{-g} (A_{3} \mp Q_{m}) F^{10}] \right|^{t = \infty}_{t = -\infty} d \theta \; d \phi \right. \nonumber \\
&+&   4 \pi a  \int^{\infty}_{-\infty} \int_{S_{\infty}} \sqrt{-g} (A^{(0)}_{3} \mp Q_{m}) j^{(2)1}  d \theta \; d \phi \; dt    \left.  \right\}
\end{eqnarray}
At $t = -\infty$, all second order terms are zero. As $t \rightarrow \infty$,
they reduce to \mbox{Kerr-Newman} type fields, with $Q^{(2)}_{m} \rightarrow 0$,
$Q^{(2)}_{e} \rightarrow \delta Q_{e}$. Therefore as $t \rightarrow \infty$,
$F^{(0)10} \propto 1/r^{2}$, $A^{(2)}_{0} \propto 1/r$;
$F^{(0)13} \propto 1/r^{4}$, $A^{(2)}_{3} \propto r^{0}$ on $S_{\infty}$, which makes the first
two  \mbox{time-integrated} terms vanish. The  \mbox{angle-integrated} term is simply $ 4 \pi (M^{2}+a^{2}) \frac{Q_{e}}{r} \delta Q_{e}$ and therefore also vanishes on $S_{\infty}$. The fourth term becomes
\begin{displaymath}
-\frac{a}{2\pi M} \int r^{2} \sin\theta \; Q_{m}(-\cos\theta)(-\frac{\delta Q_{e}}{r^{2}}) d \theta \; d \phi 
\end{displaymath}
and vanishes by  \mbox{$\theta$-integration.} Therefore
\begin{equation}
\delta (CCC)_{\rm em}=\frac{2 a}{M} \int^{\infty}_{-\infty} \int_{S_{\infty}} \sqrt{-g} (A^{(0)}_{3} \mp Q_{m}) j^{(2)1}  d \theta \; d \phi \; dt     \label{deltaCCCem}
\end{equation}
Now substituting eqs (\ref{eq:tmunu}) and (\ref{eq:j1}) into the remaining terms of $\delta(CCC)$, and combining (\ref{deltaCCCem}) with the second term of (\ref{eq:dC/dt1}),
\begin{eqnarray}  
\lefteqn{\delta(CCC) = \frac{2}{M} \int^{\infty}_{-\infty} \int_{S_{\infty}} \sqrt{-g} 
\left\{ \left[ \frac{(M^{2}+a^{2})}{8\pi}  (\partial_{0} - i e A_{0}) \psi^{*}
(\partial^{1} + i e A^{1}) \psi  \right.\right.}
\nonumber  \\
& & \left.\left. +\frac{a}{8\pi}  (\partial_{3} - i e A_{3}) \psi^{*}
(\partial^{1} + i e A^{1}) \psi 
+ \frac{M Q_{e}+a(A_{3} \mp Q_{m})}{8\pi}
ie \psi^{*} (\partial^{1} + i e A^{1}) \psi \right]  +{\rm c.c.} \right\} d \theta \; d \phi \; dt \hspace{0.8 cm}
\end{eqnarray}
where c.c. denotes the complex conjugate of the expression that precedes it and the $A$'s are all zeroth order now.  Then
we observe that $A^{1}=0$; factor out $\partial^{1} \psi$ and note that the $A_{3}$ terms cancel. So
\begin{eqnarray}
\delta(CCC) = \frac{1}{4 \pi M} \int^{\infty}_{-\infty} \int_{S_{\infty}} \sqrt{-g} 
\left\{ [ (M^{2}+a^{2}) (\partial_{0} - i e A_{0})
+ a \partial_{3} 
+ ie (M Q_{e} \mp aQ_{m})
] \psi^{*} \partial^{1} \psi \right. \nonumber \\
\left.  +{\rm c.c.}
\right\} d \theta \; d \phi \; dt \hspace{0.5 cm}
\end{eqnarray}

Since $e^{-i\omega t}$ and the monopole spheroidal harmonics $Y_{qlm}(a,\mu,\omega,\theta,\phi)$
form complete sets, we can expand the wavefunction $\psi$  in terms of the eigenfunctions:
\begin{equation}
\psi = \sum _{l,m} \int d\omega f_{lm}(\omega)
e^{-i \omega t} Y_{qlm}(a,\mu,\omega,\theta,\phi) R_{lm}(a,\mu,\omega,r)
\end{equation}
where $f_{lm}(\omega)$ are arbitrary coefficients.

Since the  integrals are taken on $S_{\infty}$, we can use the limiting form of $\psi$:
\begin{equation}
\lim_{r \rightarrow \infty} \psi  =  \sum _{l,m} \int d\omega f_{lm}(\omega)
\left\{ e^{-i \omega t}  \Theta_{qlm}(a,\mu,\omega,\theta)
                                           e^{i(m \mp e Q_{m})\phi} \frac{1}{r}
         (e^{-ikr} + A_{\omega l m} e^{ikr}) \right\}  \label{eq:exp.psi} 
\end{equation}
where the integration range is understood to exclude $-\mu < \omega < \mu$ and we have also separated the $Y_{qlm}$'s into $\theta$- and $\phi$-dependent parts.  
Substituting the above into $\delta(CCC)$, and also noting (\ref{rhosqlimit}),
\begin{eqnarray} 
 \delta(CCC)  & = &  \frac{1}{4 \pi M} \int^{\infty}_{-\infty} \int_{S_{\infty}} r^{2} \sin \theta
\sum_{l,m} \sum_{l',m'} \int d\omega d\omega '
f_{lm}(\omega) f^{*} _{l'm'}(\omega ')   \nonumber \\
&& \left\{ [ (M^{2}+a^{2})(i \omega ' -ieA_{0}) - ia(m' \mp eQ_{m})+ ie(M Q_{e} \mp aQ_{m})] 
 \right.  \nonumber \\
&&  \frac{1}{r} (e^{ik'r} + A^{*}_{\omega ' l' m'} e^{-ik'r}) 
\left[ \frac{1}{r} (-ike^{-ikr} + ikA_{\omega l m} e^{ikr})
                                                  + O(\frac{1}{r^{2}}) \right]  + \left. {\rm c.c.} \right\}  \nonumber \\
&&  \Theta_{lm}(\omega,\theta) \Theta^{*}_{l'm'}(\omega ',\theta)
e^{i (\omega '-\omega) t} e^{i(m-m') \phi}  d \theta \; d \phi \; dt
\end{eqnarray}
The contribution of the $A_{0}$ terms can be neglected, since they are one
order in $r$ lower than $\omega$ or $\omega '$; similarly the
$O(\frac{1}{r^{2}})$ terms coming from the derivatives
of the radial wavefunctions (in the square brackets) have no contribution; also the $\mp eQ_{m}$ terms cancel.

We next  take the \mbox{time-integral,} which gives a \mbox{$\delta$-function}
in $\omega$ and $\omega '$, allowing the
\mbox{$\omega '$-integral} to be done, too:
\begin{eqnarray} 
\delta(CCC) 
= \frac{1}{2\pi M}\int_{S_{\infty}} r^{2} \sin \theta
\sum_{l,m} \sum_{l',m'} \int d\omega
f_{lm}(\omega) f^{*} _{l'm'}(\omega) \Theta_{lm}(\omega,\theta) \Theta^{*}_{l'm'}(\omega,\theta)
e^{i(m-m') \phi} \nonumber \\
\left\{ [(M^{2}+a^{2}) \omega - am' + eQ_{e}M] \frac{1}{r} (e^{ikr} + A^{*}_{\omega l' m'} e^{-ikr})
\frac{k}{r} (e^{-ikr} - A_{\omega l m} e^{ikr}) + {\rm c.c.} \right\}  d \theta \; d \phi  \hspace{0.5 cm}
\end{eqnarray}

The $\phi$ and \mbox{$\theta$-integrals} can now be done (They could not be
done previously, because the orthogonality relation (\ref{eq:y.norm})
for the angular wavefunctions is only valid if the frequencies in the
functions $\Theta_{qlm}(a,\mu,\omega,\theta)$ and
$\Theta_{ql'm'}^{*}(a,\mu,\omega ',\theta)$ are the same),
giving $\delta_{ll'} \delta_{mm'}$ and thereby allowing the $l'$ and $m'$ sums to be done:
\begin{equation}
\delta(CCC) = \frac{1}{2\pi M} \int d\omega \sum_{l,m}
f_{lm}(\omega) f^{*}_{lm}(\omega)
k [(M^{2}+a^{2}) \omega - am + eQ_{e}M]
(1- A_{\omega l m} A^{*}_{\omega l m}) 
\end{equation}
Using the Wronskian relation (\ref{eq:wronski}),
\begin{eqnarray}
\delta(CCC) = \frac{1}{2\pi M} \int d\omega \sum_{l,m}
f_{lm}(\omega) f^{*}_{lm}(\omega)
[(M^{2}+a^{2}) \omega - am + eQ_{e}M] \bar{\omega} B_{\omega l m} B^{*}_{\omega l m}
 \nonumber \\
= \frac{M^{2}+a^{2}}{2\pi M} \int d\omega \sum_{l,m} f_{lm}(\omega) f^{*}_{lm}(\omega)
[\omega+\frac{eQ_{e}M-am}{M^{2}+a^{2}}]
[\omega+\frac{eQ_{e}r_{+}-am}{r_{+}^{2}+a^{2}}] B_{\omega l m} B^{*}_{\omega l m} \hspace{0.5 cm}
\label{eq:dCCCf}
\end{eqnarray}
For the extreme black hole, $r_{+} \rightarrow  M$, therefore
\begin{equation}
\delta (CCC)_{+} = \frac{M^{2}+a^{2}}{2\pi M} \int d\omega \sum_{l,m}
f_{lm}(\omega) f^{*}_{lm}(\omega)
\bar{\omega}_{+}^{2} B_{\omega l m} B^{*}_{\omega l m}.  \label{eq:dCCC+}
\end{equation}
where $\delta (CCC)_{+}$ and $\bar{\omega}_{+}$ are $\delta (CCC)$ and $\bar{\omega}$, when $r_{+} \rightarrow  M$.

Expression (\ref{eq:dCCC+}) is strictly nonnegative, but expression (\ref{eq:dCCCf}) is not. Therefore by judicious choice of frequency, $\delta(CCC)$ can be made negative for  \mbox{non-extreme} black holes, although the frequency interval we can choose from shrinks monotonically to zero as the black hole approaches extremality. Still, in contradiction to the assertions of  \cite{wang-su-et-al}, it seems that extremality can be reached by sending appropriate  \mbox{Klein-Gordon} wave packets into a black hole.

Only if all $\bar{\omega}_{+}=0$ vanish can  expression (\ref{eq:dCCC+})  be zero, therefore, the second order not dominant. Except for this very limited case\footnote{
For this case to be relevant, the wave must contain only a set of discrete frequencies, one for each $m$, if $a \neq 0$, only one frequency if $a = 0$.  In fact, the case discussed by \cite{matsas.1,matsas.2,hod.1,hod.2} fits in this category: they take $e=0$, $a=0$ and $\omega  \rightarrow  0$. Let us recall that expressions (\ref{eq:dCCCf}) and (\ref{eq:dCCC+}) are valid to second order in $\epsilon$, that is, to first order in $\delta M$, $\delta Q_{e}$ and $\delta L$. Another way to see the vanishing of $\delta(CCC)_{+}$ in \cite{matsas.1,matsas.2,hod.1,hod.2} to this order is noting that in the case they discuss, $\delta(M^2)$, $\delta(Q^2)$ and $\delta(a^2)$ all vanish, although $\delta Q$ and $\delta a$ do not vanish. Of course, it is possible to find other subcases where $\delta M$, $\delta Q$ and/or $\delta a$ do not vanish individually, but $\delta (CCC)_{+}$ still vanishes.

Then, the right-hand-side of (\ref{eq:dCCC+}) does not represent any more the contribution to lowest nonvanishing order in $\epsilon$, one has to go to higher order and/or analyze backreaction; and in this very limited case the present analysis is therefore inconclusive.
} the expression (\ref{eq:dCCC+}) for $\delta (CCC)_{+} $ will be the dominant one, and its strict positivity  means that $M^{2}$ increases faster than $(Q^{2}+a^{2})$ to the dominant (second) order in the field, meaning that {\em to this order and with the very limited, inconclusive exception noted, the Cosmic Censorship Conjecture cannot be violated by adding charge and/or angular momentum to an extreme black hole via a classical \mbox{Klein-Gordon} test field.}


\begin{thebibliography}{99}

\bibitem{penrose.orig.ccc} R. Penrose, ``Gravitational Collapse : The Role of General
      Relativity'', {\em Riv. Nuovo Cimento} {\bf 1},  special number, pp.252-276 (1969).

\bibitem{penrose.strong.ccc} Penrose, R.: Singularities and time-asymmetry, in Wald, R. M. (ed.)  General relativity: an Einstein centenary survey , Cambridge University Press, Cambridge (1979), pp.581-638.

\bibitem{ccc.rev.1} R. M. Wald, ``Gravitational Collapse and Cosmic Censorship",  \mbox{arXiv:gr-qc/9710068}

\bibitem{ccc.rev.2} R. Penrose, ``The Question of Cosmic Censorship", in  {\it Black Holes and Relativistic Stars}, ed. R. M. Wald, The University of Chicago Press, Chicago (1998), pp.103-122

\bibitem{ccc.rev.3}  P. S. Joshi, ``Cosmic Censorship: A Current Perspective", {\it Modern Physics Letters A} \textbf{17}, pp.1067-1079, (2002).

\bibitem{mazur} P.O. Mazur, ``Proof of Uniqueness of the  \mbox{Kerr-Newman}
      Black Hole Solution'', {\em J. Phys. A} {\bf15}, pp.3173-3180 (1982)

\bibitem{bunting}  G. Bunting, Ph.D. Thesis (unpublished), {\em Univ. New England} (1983).

\bibitem{nohair}  For a review, see P.O.Mazur, ``Black Hole Uniqueness Theorems" in  {\it Proceedings of the ${\it 11}^{\it th}$ International Conference on General Relativity and Gravitation}, ed. M. A. H. MacCallum, Cambridge University Press, Cambridge (1987), pp.130-157. Also  \mbox{arXiv:hep-th/0101012.}

\bibitem{wald74}R.M. Wald, ``Gedanken Experiments to Destroy a Black Hole'',
      \mbox{\em Ann. Phys.} {\bf 82}, pp.548-556  (1974).

\bibitem{semiz.1} \.{I}. Semiz, ``Dyon black holes do not violate cosmic censorship'', {\it Class. Quantum Grav.} {\bf 7}, pp.353-359 (1990).

\bibitem{hiscock}  W.A. Hiscock, ``Magnetic Charge,  Black Holes and Cosmic Censorship'',
      \mbox{\em Ann. Phys.} {\bf 131}, pp.245-268  (1981).

\bibitem{bekenst-rosenzwg} J.D. Bekenstein and C. Rosenzweig, ``Stability of the black hole horizon and the Landau ghost'',
      \mbox{\em Phys. Rev. D} {\bf 50}, pp.7239-7243  (1994).

\bibitem{hod} S. Hod,  \mbox{``Black-hole} polarization and cosmic censorship", \mbox{\em Phys. Rev. D} {\bf 60}, p.104031  (1999).

\bibitem{ford} L. H. Ford and T. A. Roman, `` `Cosmic flashing' in four
                dimensions'', {\it Phys. Rev. D} {\bf 46}, pp.1328-1339 (1992).

\bibitem{jensen} B. Jensen, ``Stability of black hole event horizons",
      \mbox{\em Phys. Rev. D} {\bf 51}, pp.5511-5516  (1995).

\bibitem{hubeny} V.E. Hubeny,``Overcharging a black hole and cosmic censorship", \mbox{\em Phys. Rev. D} {\bf 59}, p.064013  (1999).

\bibitem{hod.3} Hod,  S.: Cosmic Censorship, area theorem, and self-energy of particles, \mbox{Phys. Rev. D} 66, 024016 (2002).

\bibitem{Jacobson-Sot} Jacobson, T and Sotiriou, T.P., Over-spinning a black hole with a test body,  Phys. Rev. Lett. 103, 141101 (2009); 103, 209903(E) (2009).

\bibitem{hod-piran} S. Hod and T. Piran, ``Cosmic Censorship: The role of Quantum Gravity", \mbox{\em Gen. Rel. Grav.} {\bf 12}, pp.2333-2338  (2000).

\bibitem{deFelice-yu} F. de Felice and Yu Yunqiang, ``Turning a black hole into a naked singularity", \mbox{\em Class. Quant. Grav.} {\bf 18}, pp.1235-1244  (2001).

\bibitem{matsas.1} G. E. A. Matsas and A. R. R. da Silva, ``Overspinning a Nearly Extreme Charged Black Hole via a Quantum Tunneling Process", \mbox{\em Phys. Rev. Lett.} {\bf 99}, 181301  (2007).

\bibitem{matsas.2} G. E. A. Matsas, M. Richartz, A. Saa, A. R. R. da Silva, and D. A. T. Vanzella, ``Can quantum mechanics fool the cosmic censor?", \mbox{\em Phys. Rev. D} {\bf 79}, 101502(R) (2009).

\bibitem{hod.1} S. Hod, ``Weak Cosmic Censorship: As Strong as Ever",  \mbox{\em Phys. Rev. Lett.} {\bf 100}, 121101 (2008).

\bibitem{hod.2} S. Hod, ``Return of the quantum cosmic censor", \mbox{\em Phys. Lett. B} {\bf 668}, pp.346-349  (2008).

\bibitem{hi-dim} M. Bouhmadi-L\'{o}pez, V. Cardoso, A. Nerozzi and J. V. Rocha, "Black holes die hard: Can one spin up a black hole past extremality?", \mbox{\em Phys. Rev. D} {\bf 81}, 084051 (2010) .

\bibitem{wang-su-et-al} B. Wang, R. K. Su, P.K.N. Yu and E.C.M Young, ``Can a nonextremal  \mbox{Reissner-Nordstr\"{o}m} black hole become extremal by assimilating an infalling charged particle and shell?",
      \mbox{\em Phys. Rev. D} {\bf 57}, pp.5284-5286  (1998).

\bibitem{hawking-et-al-95} S. W. Hawking, G. T. Horowitz, and S. F. Ross, ``Entropy, area, and black hole pairs",  \mbox{\em Phys. Rev. D} {\bf 51}, pp.4302-4314 (1995).

\bibitem{t'hooft} G. `t Hooft, ``The Scattering Matrix Approach for the Quantum Black Hole: An Overview", {\em Int. J. Mod. Phys. A} {\bf11}, pp.4623-4688 (1996).

\bibitem{numerical} S. L. Shapiro and S. A. Teukolsky, ``Formation of naked singularities: The violation of cosmic censorship", {\it Phys. Rev. Lett.} {\bf 66}, pp.994-997 (1991).

\bibitem{mtw} C. W. Misner, K. S. Thorne and J. A. Wheeler, {\it Gravitation},
      Freeman, New York (1973).

\bibitem{wu-yang} T. T. Wu and C. N. Yang, ``Dirac monopole without strings: Monopole harmonics",
                                   {\it Nucl. Phys. B} {\bf 107}, pp.365-380 (1976).

\bibitem{semiz.2} \.{I}. Semiz, ``Klein-Gordon equation is separable on the
      dyon black-hole metric'', {\it Phys. Rev. D} {\bf 45}, pp.532-533 (1992). Erratum: {\it Phys. Rev. D} {\bf 47}, p. 5615 (1993).

\end{thebibliography}
\end{document}